
\magnification 1100
\rightline{SPhT 95/058}
\rightline{DAMTP 95/25}
\vskip 1.5 truecm
\centerline{\bf THE COULOMB GAS BEHAVIOUR}
\centerline{\bf  OF TWO DIMENSIONAL TURBULENCE}
\vskip 2 cm
\centerline{{\bf Ph. Brax}\footnote{$^1$}{On Leave of Absence from
SPhT, CE-Saclay, F-91191 Gif-sur-Yvette cedex, France}
\footnote{$^@$}{P.Brax@amtp.cam.ac.uk}}
\vskip .3 cm
\centerline{Department of Applied Mathematics and Theoretical Physics}
\centerline{Silver Street, Cambridge, CB3 9EW, England}
\vskip .3 cm
\vskip 3 cm
\centerline{ABSTRACT}
\vskip .5 cm
The long-time large-distance behaviour of free decaying two
dimensional turbulence is studied. Stochastic solutions of
the Navier-Stokes equation are explicitly shown to follow renormalisation
group trajectories.
 It is proven that solutions of
the Navier-Stokes equation asymptotically converge  to fixed points which
are conformal field theories. A particular fixed point
is given by the free Gaussian field with a charge at infinity.
The stream function is
identified with a vertex operator. It happens that this solution also
admits  constant $n$-enstrophy fluxes in the asymptotic regime,
therefore fulfilling all requirements to represent an
asymptotic state of two-dimensional turbulence.
 The renormalisation basin of attraction of this
fixed point consists of a charged Coulomb gas. This Coulomb gas gives
an effective description of turbulence.
\vfill \eject
\vskip 1 cm
\leftline{\bf I Introduction}
\vskip .5 cm
Two-dimensional turbulence is a simplified version of
three-dimensional
 turbulence. It can nonetheless give qualitative
information on phenomena such as atmospheric turbulence.
Recently, many field theorists have  been attracted by 2d turbulence.
 This new interest has been aroused by
Polyakov's seminal$^{[1]}$ paper. For the first time, it seems
 that this long-standing problem
can be tackled using field theoretic methods$^{[2]}$.
Definite predictions are obtained. One can hope to
compare them with numerical simulations$^{[3]}$.

Most analyses of two-dimensional turbulence aim at
  computing the two-dimensional analogue
of
the celebrated Kolmogorov spectrum$^{[4]}$in the long wavelength
limit
. An early analysis by
Kraichnan$^{[5]}$ showed that the energy spectrum should decrease as $k^{-3}$,
due to the transfer of vorticity from large to small scales where
dissipation takes place. Nevertheless, this prediction suffers from large
infrared divergences and is corrected by logarithms at one loop order.
On the other hand, such a correction does not assure convergence at two
loop
order where double logarithm infrared divergences appear.
Alternative predictions have been given by Saffman$^{[6]}$ and
 Moffat$^{[7]}$ with respective spectra $k^{-4}$ and $k^{-{11\over
3}}$.
This variety of results shows that the situation is far from being
settled. From a field theoretical point of view, these approaches
have severe drawbacks, they all assume that dimensional analysis is
exact and neglect possible non-trivial anomalous dimensions appearing
in short distance expansions of field products. Unfortunately,  no
precise rationale can be extracted from numerical
simulations.
 It happens that simulations starting
from gaussian initial conditions show an increase of the
exponent from -4 to -3 for short times. Then coherent vortices
emerge. Moreover, a very long time
simulation$^{[8]}$ shows that the fluid motion seems to converge to
 an ensemble of point vortices$^{[9]}$. The fate of the spectrum
exponent as time goes to infinity is certainly not clear.

The Polyakov  approach  deals with steady
states of two-dimensional turbulence.
 This amounts to imposing boundary conditions and an external
pumping force. Turbulence is then only obtained in the inviscid limit
where
the viscosity goes to zero. Interpreting the Navier-stokes equation
 as
an equation for correlation functions, Polyakov shows that
scale-invariant
 field theories can represent steady states if a certain
inequality
 is satisfied by
operator dimensions. This boils down to imposing
a vanishing non-linear term in the Navier-Stokes equation. Another
constraint is obtained by forcing the enstrophy flux to be a constant. A
simplified treatment of the resulting Diophantine equation leads to a
solution whose spectrum scales as $k^{-25/7}$. Unfortunately, such a steady
state is non-unitary and thus suffers from large infrared problems.
The Diophantine equation can be shown to yield thousands of possible
non-unitary minimal conformal field theories$^{[10]}$. As emphasised
in ref.[11], these solutions are hardly physical. In addition to
infrared divergences, they fail to provide a non-zero enstrophy flux.
Moreover, they only take into account the enstrophy flux and discard
fluxes of higher conserved quantities. It has been recently shown
$^{[12]}$ that a conformal description of two dimensional steady states
is rather unlikely.

In this paper, we strive to tackle a less ambitious problem, i.e
free decaying two dimensional turbulence. To do so, we follow closely
the Polyakov approach but try to remedy some of its drawbacks by imposing
that fluxes of all conserved quantities are constant.
As time increases, viscosity effects become less important and
solutions of the Navier-Stokes equation reach a quasi-equilibrium
containing free coherent vortices. For such a state, it seems
reasonable to expect that fluxes  as well as the energy are
zero. The object of this article is to show that these conclusions
can be derived from an effective field theory describing the long-time
 regime of two-dimensional turbulence.

The paper is arranged as follows. In a first part, some comments on
the nature of turbulence are proposed.
The second part deals with solutions of Navier-Stokes
equations.
In the third part, we impose that fluxes of conserved quantities are
 constant. Then, the long time behaviour of conserved charges is
analysed. In the last section, we propose an effective theory for two
dimensional free decaying turbulence.
\vskip 1 cm
\leftline{\bf II A Field Theoretical View of Turbulence}
\vskip 1 cm

Turbulence is often described as a random behaviour reached by fluids
under certain conditions such as high speed and low viscosity. Fluid
dynamics describe any fluid motion in terms of the velocity field $v$.
This field is a $macroscopic$ field: it is the average of molecular
velocities over a macroscopic size $a$ ( $a$ is rather small and can
be thought to be of the order of a micron). Below this scale, the
Hydrodynamic approximation is not valid. There is another intrinsic
scale $l$; it is set by viscosity effects and represents the scale
below which viscosity smoothes variations of the velocity field. Above
this scale, the velocity is rapidly varying.
 A sufficiently low viscosity is a phenomenological
condition for the existence of a turbulent regime.
 In that case, the viscous length $l$ becomes of
the order of the macroscopic length $a$. We suppose that this
criterion is fulfilled, so that there is a single fundamental length scale $a$.

The equation governing the time evolution of the velocity field is a
 non-linear equation,
$$(\partial_t +v.\partial)v=\nu \Delta v -\nabla p+F\eqno(1)$$
known as the Navier-Stokes
 equation where $p$ is the pressure field  and $F$ the
external pumping force.
Such an equation is  simpler in two dimensions as it only
 involves a pseudo-scalar field instead of a vector field.
This allows  strong results about classical solutions$^{[12]}$ to be proven.
It can be shown  that smooth (${\cal C}^{\infty}$)
initial conditions remain smooth.
However, this is not the sort of statement we are interested in. Indeed,
we would like to understand the Navier Stokes
 equation when the velocity field is
far from being smooth.

As a matter of fact, the velocity field is a {\bf random field}. This
means that each field configuration is chosen at random according to a
probability law. We will restrict ourselves to probability laws which
can be expressed as  statistical field theories. Finding the
probability law amounts to finding the action $S_t$ giving the Boltzmann
weight at every time $t$.
The action $S_t$ is an {\bf effective action} valid for distances
larger than the macroscopic length $a$. It is a local action.
Indeed, the range of the interaction between two fluid elements
(by this we mean a macroscopic neighbourhood of a point) is
given by the viscosity length $l$. Below this scale, the velocity field
hardly varies due to viscosity effects; on the contrary, for larger
scales, two fluid elements are not directly in contact, and their velocity
fields can be almost uncorrelated. But for turbulent fluids,  the viscous
length is of the order of the macroscopic length. The length $a$ plays
the role of a physical Ultra-Violet cut-off, so that the interaction
length is of the order of the UV cut-off. As all fields will be
smeared out (regularised) over a distance $a$, this implies that the
Lagrangian composing the action is in fact local, i.e. can be expressed
as a function of regularised fields and their derivatives. This does
 not prevent the existence
of any long-range order. As for critical phenomena, long-range order
 can occur due
to collective effects even though interactions are short-ranged.

Two highly different  issues could be tackled. On the one
hand, the  Ultra-Violet problem  amounts
to solving the Navier-Stokes equation with an external pumping force and
given boundary conditions. Most of the time, a large box and a low
frequency random pumping force are specified.
 Turbulence  only appears in the inviscid limit where $\nu$ goes to zero.
 The velocity field is random because of the random external force.
Notice that in field theoretic parlance, this is tantamount to studying
the UV behaviour of the action $S_t$, keeping an Infra-Red cut off
fixed. Scale invariant solutions are then time-independent UV
 fixed points in the
renormalisation group sense. They represent stationary turbulence
states.
On the other hand, the Infra-Red problem deals with infinite space and
Dirichlet boundary conditions at infinity. No external pumping is
required but  random initial conditions are imposed. As shown in
section III, the viscosity $\nu$ is finite and does not need to be
scaled down to zero. It becomes negligible in the long time regime.
This is a free decaying turbulence situation where scale invariance
is gradually restored by time evolution. It is quite analogous to a
self-criticality regime. This issue has to be faced as an Infra-Red
problem, where long-time, large-scale limits are taken, keeping the UV
cut-off $a$ fixed. Asymptotic solutions are then IR fixed points.
The relation between IR and UV fixed points seems to be a difficult
issue.

 It is convenient to move from the
velocity picture to the vorticity picture.
 In 2d, the continuity
equation reads
$$\nabla v =0,\eqno(2)$$
valid for an incompressible fluid. This entails that $v$ is a curl:
$$v_{\alpha}=e_{\alpha\beta}\partial_{\beta}\psi\eqno(3)$$
where $\psi$ is the stream function.
The field $\psi$ is a pseudo-scalar field.
In the plane, it is always possible to construct  pseudo-scalar fields
 from scalar fields,
they are just two-forms $\psi dz\wedge d\bar z$.
It is also useful to introduce the
  vorticity which is  the curl of $v$:
$$\omega =e_{\alpha\beta}\partial_{\alpha}v_{\beta}.\eqno(4)$$
Notice that $e_{12}=-e_{21}=1$ is the antisymmetric tensor, i.e.
$\omega =-\Delta \psi$.
This is also a pseudo-scalar field.
 This yields a different form of the Navier-Stokes
equation:
$$\partial_{t}\omega +e_{\alpha\beta}\partial_{\alpha}\psi
\partial_{\beta}\Delta \psi =\nu \Delta \omega\eqno(5)$$
This equation for the vorticity field will be the main ingredient of
our analysis.

Before further delving into the properties of the vorticity field, we
will  recall a few facts about certain two-dimensional field theories.
The velocity field  obeys the Navier-Stokes equation.
It is not clear what meaning can be given to the product of fields
appearing in the non-linear term. The following will provide an answer.
Fields will be chosen to be  tempered distributions $\psi \in {\cal S}'$
(continuous linear forms on the space of rapidly
decreasing functions $\cal S$). A field theory such as the IR fixed
point is a measure on the space of tempered distributions$^{[13]}$.
To regularise the short distance behaviour of fields, introduce a
smoothing function $\rho_{\epsilon}$:
$$\eqalign{
\int d^2x \ \rho (x)=1\cr
\rho_{\epsilon}(x)={1\over \epsilon^2}\rho ({x\over \epsilon})\cr
\rho\ge 0.\cr}
\eqno(7)$$
This function is an approximate delta function. It allows to transform
the very rough variations of tempered distributions into perfectly
well behaved functions. To do so, introduce the {\bf regularised field}
$$\psi_{\epsilon}(x)=
\int d^2y\  \psi(y)\rho_{\epsilon}(y-x)\eqno(8)$$
This is now a smooth (${\cal C}^{\infty}$) function. We will interpret
 $macroscopic$ fields as regularised fields where $\epsilon=a$ the
physical UV cut-off. This allows to see the Navier-Stokes equation and its
vorticity counterpart as equations for regularised fields. The non-linear
term is then a simple product of smooth functions. Obviously care will
be needed in removing the cut-off.

 The Navier-Stokes equation is now expressed in terms of regularised
fields. Physically, this is all the more natural as the velocity field is
the average of molecular velocities. This gives:
$$\partial_t \omega_a +e_{\alpha\beta}\partial_{\alpha}\omega_{a}
\partial_{\beta}\Delta \psi_{a}=\nu \Delta \omega_a.\eqno(9)$$
The initial time field theory is specified down to the cut-off $a$.
It is the initial state of the Cauchy problem for the evolution
equation. It is specified by an initial effective field theory
representing the initial probability law of the stream function. At
any given time $t$,
the effective field theory describing the behaviour of $\psi_a$ is
obtained from (9). It represents the probability law describing the
fluctuations of $\psi_a$ at time $t$.

Suppose that the effects of viscosity can be neglected
 in the asymptotic regime. Then the inviscid fluid
possesses an infinite number of conserved charges. The first one is
obvious, it is the energy per unit mass. Another set of conserved
quantities stems from the pseudo-scalar behaviour of the
vorticity. Introduce the generalised n-enstrophy:
$$H_{n,a}=\int d^2 x\  \omega_a^n(x).\eqno(10)$$
Only the even enstrophies are non zero.
Now compute
$${{dH_{n,a}}\over dt}=ne_{\alpha\beta}\int d^2x\
\partial_{\alpha}\psi_{a}\partial_{\beta}\Delta \psi_{a}
\ \omega_{a}^{n-1}(x).\eqno(11)$$
Integrate by parts using the antisymmetry of $e_{\alpha\beta}$
to obtain:
$${{dH_{n,a}}\over dt}=e_{\alpha\beta}
\int  d^2x\
\omega_{a}^n \partial_{\alpha}\partial_{\beta}\Delta\omega_{a}
\equiv 0.\eqno(12)$$
We will need to integrate by parts very frequently. So from
now on, we restrict  fields to be zero at infinity, ie
 regularisations $\psi_a,\ \omega_a...$
 vanish at infinity. The second moment is simply called the Enstrophy.
In section III, we will justify this calculation directly from the
viscous Navier-Stokes equation. It will be shown that this result is
indeed true  in the long-time regime, even in the presence of viscosity.
Such conserved charges give rise to conserved currents in Fourier
space.
These currents represent the flux of n-enstrophies from  scale $q$
to $q+dq$. In the usual Kolmogorov approach, the energy flux (called
the energy transfer rate $\epsilon$) is supposed to be finite and
constant. Physically, these conditions imply that every point in the
fluid plays the role of a local source of n-enstrophy and energy.
 In the free decaying case, we
expect these fluxes to be zero if a steady state exists. In the
following we will call turbulence any critical solution of Navier-Stokes
equations such that the associated fluxes are $finite$ and $constant$
(possibly zero).
\vskip 1 cm
\vfill \eject
\leftline{\bf III General Asymptotic Properties Of Decaying
Turbulence}
\vskip 1 cm
\leftline{\bf a) The Neighbourhood of Renormalisation Group Fixed
Points}
\vskip .5 cm
 We will suppose that turbulence is specified by a massless field theory. This
theory is a fixed point of the evolution equations in the long
time-large scale regime. From  given initial conditions, it is not
 clear what the set of fixed points can be. It could be
 a single theory, and in that case any initial condition
would finally be  attracted by this fixed point. It could also be a more
complicated set such as a manifold.

On physical grounds, 2d turbulence  is likely to be
represented by a unique fixed point. Indeed,
2d turbulence is isotropic. Then, universality suggests that there
must be a unique
universality class describing 2d turbulence in the infrared regime.
This is certainly true if the notion of universality can be extended to
self-organised
criticality.
 In the following, we will try to first analyse the
features of  a fixed point. Thus, we will show that any conformal
theory satisfying a specified inequality can be a fixed point. Fixed
points are
actually  asymptotic solutions of Navier-Stokes if they are  stable
fixed points, i.e. if nearby deformations of a fixed point converge
in the renormalisation group sense.

First observe that  the viscosity defines a length scale and
a time scale, i.e. by engineering dimensional analysis:
$$\nu={{a^2}\over \tau}.\eqno(13)$$
The scale $a$ is the UV cut-off whereas typical
times are measured in units of $\tau$. We want to investigate the
large distance $r\gg a$ long time $t\gg \tau$ regime.
Let us first define a fixed point as a continuum conformal field
theory satisfying the Euler equation:
$$\mathrel{\mathop {\lim}_{\lambda\to\infty}}[{{d\omega_{a\over
\lambda}}\over {dt}}+e_{\alpha
\beta}\partial_{\alpha}\psi_{a\over\lambda}\partial_{\beta}\Delta\psi_{a\over
\lambda}]=0.\eqno(14)$$
Notice that such an equation involves the product at shorter and
shorter distances of two regularised fields. In
order to further analyse the Euler equation, we
need to evaluate short distance expansions.
 In a nutshell, the expansion
$$\psi (y) \phi (z)=
\sum_{\alpha}C_{\alpha}(y,z){\cal O}_{\alpha}(z)
\eqno(15)$$
is valid in any correlation function.
The coefficient functions are in fact distributions which
scale as $v=\vert y-z\vert\to 0$ as:
$$C_{\alpha}(y,z)\sim v^{d_{{\cal O}_{\alpha}}-d_{\psi}-d_{\phi}.}\eqno(16)$$
Moreover, the most singular term on the right-hand side can be retained
as the most significant piece.
The Euler equation  involves a non-linear term whose continuum limit
is far from being obvious. Nevertheless, it can be computed using the
operator product of two $\psi$.
Indeed, the non-linear term reads (we put $\epsilon={a\over \lambda}$)
$$e_{\alpha\beta}\partial_{\alpha}\psi_{\epsilon}(x)\partial_{\beta}
\Delta \psi_{\epsilon}(x)=\int d^2u d^2v e_{\alpha\beta}
\partial_{\alpha}\rho_{\epsilon}(u-x)\partial_{\beta}\Delta
\rho_{\epsilon}(v-x)\psi(u)\psi(v).\eqno(17)$$
In the continuum limit $\epsilon \to 0$, it is safe to  retain the dominant
term in the operator product expansion:
$$\int d^2ud^2ve_{\alpha\beta}
\rho_{\epsilon}(u-x))\rho_{\epsilon}(v-x)
\partial^{u}_{\alpha}
\partial^v_{\beta}\Delta^v\sum_{\gamma}
C_{\gamma}(u,v)\psi_{2}^{\gamma}(v),\eqno(18) $$
where $\psi_2^{\gamma}$ are the fields appearing in the short distance
expansion.
The only way of getting a non-zero term is to act on the possibly
non-symmetric term under $\alpha\to\beta$. One of the coefficient
functions
 must satisfy
$C(u_1,u_2,v_1,v_2)=-C(v_2,v_1,u_2,u_1)$.
This can only be obtained from descendent terms, where the leading
operator is constructed from a pseudo-scalar combination of descendent
fields $L_{-n1}....L_{-n_m}\psi$. The Virasoro generators $L_n={1\over
2\pi}
\int dz
z^{n+1}T(z)$ are the generators of the 2-d conformal algebra and $T$
is the energy momentum tensor.
 Truncating the series yields:
$$\int d^2v d^2u \rho_{\epsilon}(v)\rho_{\epsilon}(u-x)
 e_{\alpha\beta}\partial^v_{\alpha}\partial^u_{\beta}
\Delta^uC_{\gamma}(u+v,u)\psi_2(u)\eqno(19)$$
After changing variable $v\to \epsilon v$, and using the fact that the
cut-off is small:
$$\int d^2v \rho(v) e_{\alpha,\beta}\partial_{\alpha}^v
\partial_{\beta}^u\Delta^u C_{\gamma}(u+\epsilon v,u)\sim
\epsilon^{d_{\psi_2^{\gamma}}-2d_{\psi}-4}\int d^2v
e_{\alpha\beta}\partial_{\alpha}^v\partial_{\beta}^u\Delta^u
C_{\gamma}(u+v,u).\eqno(20)$$
 We will denote the constant by C, neglecting its
dependence on $u$:
$$e_{\alpha\beta}\partial_{\alpha}\psi_{\epsilon}\partial_{\beta}
\Delta \psi_{\epsilon}\sim \epsilon^{d_{\psi^{\gamma}_2}-
2d_{\psi}-4}
C\psi_{2,\epsilon}^{\gamma}(x).\eqno(21)$$
{}From now on, all fields will be normalised in such a way that finite
constants like $C$ are equal to one.
It is now necessary to identify the leading operator. Notice that due
to the term $e_{\alpha\beta}$ the result  is a pseudo-scalar.
Futhermore it has to be antisymmetric
under $u_1\to u_2$.
 Using complex coordinates:
$$\eqalign{
\psi_2^{\gamma}(i\bar z, iz)=-\psi_{2}^{\gamma}(z,\bar z)\cr
\psi_2^{\gamma}(\bar z,z)=-\psi_2^{\gamma}(z,\bar z)\cr}
\eqno(22)$$
Under the first transformation the energy momentum tensor
transforms as:
$$\eqalign{
T\to -\bar T\cr
\bar T\to -T\cr}\eqno(23)$$
This yields the transformations of Virasoro generators
 $L_{n}\to i^n \bar L_n$.
The lowest order
anti-invariant  combination is:
$$\psi_2^{\gamma}=(L_{-2}\bar L_{-1}^2-\bar L_{-2}L_{-1}^2)\psi_2.\eqno(24)$$
This immediately gives the leading behaviour in (20) as
$$d_{\psi_{2}^{\gamma}}=d_{\psi_2}+4.\eqno(25)$$
This calculation  can be extended by induction to higher order short distance
expansions.
After this lengthy computation, we obtain the conditions for the
existence of a fixed point. Suppose that the fixed point theory is time
invariant. Then the Euler equation is satisfied if the non-linear term
vanishes altogether in the continuum limit. This implies that$^{[1]}$:
$$d_{\psi_2}>2d_{\psi}\eqno(26)$$
This is the Polyakov inequality.
Hence, all conformal field theories satisfying this equation are
solutions of the Euler equation. However, they may not be valid representation
of 2d turbulence. To select the expected unique fixed point modelling
2d turbulence, we will have to impose that fluxes of $n$-enstrophies are
constant in the infra-red regime.

We are now in position to prove that fixed points are asymptotic
solutions of the Navier-Stokes equation. Suppose that, after
a transient period of order $O(\tau)$, the field theory obtained solving
the Cauchy problem from a given initial condition falls within the
renormalisation group neighbourhood of a fixed point (we will call
this neighbourhood the basin of attraction of the fixed point). This means that
this field theory converges in the large distance limit towards the
fixed point.  Let us prove that such
renormalisation group trajectories coincide with the Navier-Stokes
evolution. Suppose that  large distance long time trajectories are
renormalisation group trajectories. To implement this idea, rescale
$x\to \lambda x$ and $t\to \lambda^T t$ where $T$ is a yet unknown constant.
Then  define a new field $\tilde\psi$  as the regularised field
following the renormalisation trajectories of the solution $\psi$ at
time $t$.
This means that the regularised field
$\tilde\psi_{a}(\lambda x,\lambda^T t)$
obeys the following:
$$\tilde\psi_{a}(\lambda x,\lambda^T t)=
\lambda^{-d_{\psi}}\psi_{\lambda^{-1}a}(x, t).\eqno(27)$$
The field $\tilde \psi$ is the effective field obtained after
integrating the action $S_{t}$ over distances in the range
$[\lambda^{-1}a,a]$, whereas the field $\psi_{\lambda^{-1}a}$ is the
regularised field defined with the action $S_{t}$.
Notice that this ansatz $\tilde\psi$ is a solution of Navier-Stokes
equation provided that, when plugging  $\tilde\psi$ in (9), one ends
up with the original the Navier-Stokes equation for $\psi$ at time $t$ and
cut-off $a\over \lambda$. Obviously, such an equation should have a
modified viscosity $\nu \to {\nu\over\lambda^{2-T}}$ as the typical
time $\tau$ is also rescaled to $\tau\over\lambda^{T}$.
Putting this ansatz back in Navier-Stokes equation after adjusting the
 time rescaling $t\to
\lambda^{2+d_{\psi}}t$, one obtains
$$\partial_t \omega_{\lambda^{-1}a}+e_{\alpha\beta}
\partial_{\alpha}\psi_{\lambda^{-1}a}\partial_{\beta}\Delta \psi_{
\lambda^{-1}a}=\lambda^{-2+T}\nu\Delta \omega_{\lambda^{-1}a}.
\eqno(28)$$
This is indeed the expected result. This proves that $\tilde\psi$ is a
solution.
Notice that scale invariance is asymptotically achieved if and only if
the dimension $d_{\psi}$ is negative. Moreover $T$ has to be positive
in order to ensure that time flows towards infinity. This implies that
$-2<d_{\psi}<0$. Hence, we retrieve Polyakov's idea that
turbulence requires non-unitary theories. This entails an extreme
boundary condition sensitivity. We will end up
considering charges at infinity which in a sense epitomise the
non-trivial
 infrared properties of turbulence.
Notice  that viscosity explicitly breaks scale invariance.
 The effective action in
the basin of attraction of the fixed point will reflect this fact, i.e.
it will be the sum of the fixed point action and irrelevant operators
flowing to zero in the infrared regime.
To conclude,
the field $\psi_{a\lambda^{-1}}$ tends to the continuum limit field
$\psi$ whereas the solution $\tilde\psi$
converges to the fixed point along renormalisation group trajectories.
This proves that fixed points are asymptotic solutions of
Navier-Stokes equation
\footnote{$^1$}{This result is also valid in three dimensions. In that
case, solutions of the Navier-Stokes equation still follow renormalisation
group trajectories. In the infrared limit, solutions converge to
non-unitary conformal fixed points defined by a straightforward
generalisation of (14). On the contrary, the Polyakov inequality seems to
require a non-trivial extension.}  . Amongst all these fixed points, we will
explicitly find an example admitting finite fluxes in the infra-red regime.
\vskip .5 cm
\leftline{\bf c) The Effects of Viscosity on $N$-Enstrophies}
\vskip .5 cm
As viscosity is introduced, it is no longer true that $n$-enstrophies
are conserved. Nevertheless, they are slowly varying.
In order to assess the effect of viscosity, let us compute
the $n$-enstrophies and their time derivatives in the asymptotic
regime. By (10),
$$H_{n,a}(\lambda^T t)=\int d^2x\ \omega_{a}^n(x,\lambda^T t),\eqno(29)$$
Change $x\to \lambda x$ and use the scaling properties of $\omega$ to obtain
$$H_{n,a}(\lambda^T t)=\lambda^{2-2n-nd_{\psi}}\int d^2x\
\omega_{\lambda^{-1}a}^n(x,t).\eqno(30)$$
The integrand tends to be the product of $n$ fields at the same point,
so  it is necessary to evaluate higher order short distance
expansions. These products are defined by induction.
They can be written in a compact way using conformal families:
$$[\psi][\psi_{n-1}]=[\psi_{n}]+....\eqno(31)$$
where $\psi_n\equiv\hbox{Dom}(\psi\psi_{n-1})$ is the leading part of
the short distance expansion $\psi\psi_{n-1}$.
This entails that inductively:
$$\omega_{\lambda^{-1}a}^n(x,t)\mathrel{\mathop{\sim}_{\lambda\gg 1}}  (\lambda
a^{-1})^{-d_{\psi_n}+nd_{\psi}+
2n}  \psi_{n,\lambda^{-1}a}(x,t).\eqno(32)$$
Hence the $n$-enstrophy:
$$H_{n,a}(\lambda^T t)=
\lambda^{2-d_{\psi_n}}a^{d_{\psi_n}-nd_{\psi}-2n}
\int d^2x \ \psi_{n,\lambda^{-1}a}(x,t).
\eqno(33)$$
$N$-Enstrophies are simply integrals of higher order fields.

Similarly, the time derivative reads
$${{dH_{n,a}(\lambda^T t)}\over dt}=n\lambda^{2-2n-nd_{\psi}}
\int d^2x\  {\nu \lambda^{d_{\psi}}}\Delta\omega_{\lambda^{-1}a}(x,t)
\omega_{\lambda^{-1}a}^{n-1}(x,t)\eqno(34)$$
where the rescaled version (28) of Navier-Stokes equation is used.
The right-hand side is always negative as integration by parts
yields $-(n-1)\nu\int d^2x \ (\partial \omega_{\lambda^{-1}a})^2\omega_{
\lambda^{-1}a}^{n-2}$.
Finally,
$${{dH_{n,a}(\lambda^Tt)}\over dt}=-n \nu  \lambda^{4+d_{\psi}-d_{\psi_{n}}}
a^{-2(n+1)-nd_{\psi}+d_{\psi_n}}
\int d^2x\  \psi_{n,\lambda^{-1}a}(x,t).\eqno(35)$$
Using the scaling properties of the fields, this yields in the
infrared regime,
$${{dH_{n,a}(t)}\over dt}=-n{\nu\over a^2}  H_{n,a}(t)
\eqno(36)$$
where $t$ is now the $\ asymptotic\ physical\ time$. Notice that the transient
time
${1\over \tau}$ appears.
The $n$-enstrophies go to zero in the asymptotic regime.
Similarly, one gets
$$E_a (t)=E_a(\tau)\exp(-2{ {t-\tau}\over \tau})\eqno(37)$$
in the asymptotic regime.
Notice that the energy is exponentially damped to zero unless there is no
viscosity.

So we have found that in the infrared regime solutions of
Navier-Stokes equation have vanishing conserved quantities. This does not
imply however that solutions are trivial.
\vskip .5 cm
\leftline{\bf c) The Energy Spectrum}
\vskip .5 cm
 In the same vein, one can obtain an expression for the scaling
exponent of the energy spectrum.
The spectrum  is nothing but the Fourier transform of the velocity connected
2-point function:
$$E(k)=2\pi k \int d^2x \exp (-i2\pi k.x)<v_{a,\alpha}(0)v_{a,\alpha}(x)>
\eqno(38)$$
such that
$$E_a=\int dk E(k)\eqno(39)$$
The scaling behaviour of $E(k)$ in the inertial range can be deduced as
follows:
$$<\partial_{\alpha}\psi_a(\lambda x,\lambda^T
t)\partial_{\beta}\psi_a(\lambda y,\lambda^T t)>
\sim \lambda^{-2-2d_{\psi}}<\partial_{\alpha}
\psi_{\lambda^{-1}a}(x,t)\partial_{\alpha}\psi_{\lambda^{-1}a}(y,t)>.
\eqno(40)$$
The two point function can be evaluated in the continuum limit as
$\lambda$ goes to infinity:
$$<\partial_{\alpha}\psi_{a}(\lambda x,\lambda^T t)\partial_{\alpha}
\psi_a(\lambda y,\lambda^T t)>\sim (\lambda x-\lambda y)^{-2-
2d_{\psi}}.\eqno(41)$$
Putting $y=0$ yields the large scale behaviour of the velocity two
point function.
The Fourier transform scales  as
$$E(k)\mathrel{\mathop{\sim}_{k\to 0}} k^{2d_{\psi}+1}.\eqno(42)$$
Notice that this is the {\bf infra-red } spectrum. Once a fixed point
is specified, the infrared spectrum follows. As expected, no
prediction on the UV spectrum is derived.
\vskip .5 cm
\leftline{\bf d) Some Examples}
\vskip .5cm
 Let us now give the simplest possible solution of
the Navier-Stokes equation: the Gaussian free field with charges at
infinity$^{[16]}$.
 Choose the Lagrangian
to be:
$$S={1\over 4\pi}\int d^2x\ (\partial\phi)^2\eqno(43)$$
 and identify $\psi=:\exp i\phi:$.
We explicitly suppose that there is a charge $Q$ at infinity.
This conformal field theory can be abstractly defined by its
energy-momentum tensor and the operator product algebra. One can also
use also a formal device to write down the action of the deformed
Gaussian theory with a charge at infinity. Indeed, it is sufficient to
add to the free Gaussian action a source term  $\int d^2x
J_{\infty}(x)\phi$ where the source is $J_{\infty}=iQ\delta^2(x-\infty)$.
 This formal source at infinity
implies that correlation functions of vertex operators are non-zero
provided the sum of the exponents of each operator is equal to $Q$.
  Then the field $\psi_2=:\exp 2i\phi:$
has dimension $2+Q$ whereas $d_{\psi}=0.5(1+Q)$. So vertex operators are good
candidates to represent the asymptotic regime of the stream
function. The charge at infinity is restricted to $-5\le Q<-1$.
This theory is quite peculiar. Indeed, all UV properties, such as
short distance expansions, are highly dependent on the value of the
charge at infinity. In a sense, we can say that this reflects the
non-unitarity of the theory and the increase of correlation functions
with distance, i.e. large distance properties influence small scale
characteristics.

We would like to interpret this theory in a probabilistic way where
operators are replaced by fields. This require that we identify
$:\exp i\phi:$ with its formal adjoint $:\exp(-(1+Q)i\phi):$.
Therefore, we deduce that\footnote{$^2$}{The central charge is $c=-11$}:
$$Q=-2.\eqno(44)$$
Notice that in this case the infrared spectrum behaves as
$$E\mathrel{\mathop{\sim}_{k\to 0}}k^{0},\eqno(45)$$
and in fact it  happens  this apparently innocuous solution is also
turbulent.
 This
demonstrates the idea that two dimensional turbulence is a problem where large
scales are relevant.

\vskip 1 cm
\leftline{\bf IV- Turbulent Fixed Points}
\vskip 1 cm
\leftline{\bf a) Currents in the Infra-Red Limit}
\vskip .5 cm
The conventional picture of turbulence requires that some
physical entity (eg the energy) is transferred from large to small
scales. This means that currents are non zero constants. We have
proven that in the continuum limit $n$-enstrophies and the energy are
conserved quantities.
The conserved quantities can be  written down as  reciprocal-space integrals:
$$H_{n,a}=\int dq\  h_{n,a}(q)\eqno(46)$$
The right-hand side is only modulus dependent. The integrand represents the
conserved quantity per unit mode.  As $H_{n,a}$ is asymptotically
conserved, there is a conserved current as well.
More precisely,
$${{dh_{n,a}}\over dt}+\partial_q R_{n,a}(q)=0\eqno(47)$$
where $R_{n,a}$ is the current.
The current involves the Fourier transform of
$\omega_a^n(x)$
and reads
$$\eqalign{
h_{n,a}(q)=2\pi q
L(-q)(\omega_{a}^{n})(q)\cr
R_{n,a}(q)={d\over dt}\int_{k>q} h_{n,a}(k)\cr}
\eqno(48)$$
where  $L(q)$ is the Fourier
transform of the characteristic function of  a bounded domain of
integration $L$.
The next step is to come back to real space.
Introduce the structure factor
$$\theta_q(x)=\int_{k>q} d^2k\  e^{i2\pi k.x}.\eqno(49)$$
Then
$$R_{n,a}={d\over dt}\int_{L} d^2x\  \theta_q *
\omega_{a}^{n}.\eqno(50)$$
The infinite volume limit can now be taken.
Notice that the currents  can be written
as integrals of  densities:
$$\eqalign{
R_{n,a}(q)=\int d^2 x \ r_{n,a}(x)\cr
r_{n,a}(x)=\theta_q*{{d}\over {dt}}\omega_{a}^n(x)\cr}
\eqno(51)$$
The time-derivative is explicitly taken as in the Euler equation.
Currents appear as  integrals of  current densities which
depend on $q$. In the Kolmogorov theory, current densities are
constant and finite.
We would like to evaluate  current densities in the asymptotic regime.
So rescale space and time
such that
$$r_{n,a}(\lambda x,\lambda^T t,q)=\lambda^{-d}r_{n,\lambda^{-1}a}
(x,t)\eqno(52)$$
where $d$ is the dimension of $r$. The current density is given by a
convolution; it can be simplified by noticing the scaling property:
$$\theta_q(x-y)=\lambda^2 \theta_{q\over \lambda}
(\lambda x-\lambda y).\eqno(53)$$
Then
$$r_{n,\lambda^{-1}a}(x,t,q)=
-n \int d^2y \lambda^2 \theta_{q\over \lambda}(\lambda x-\lambda y)
e_{\alpha \beta} \partial_{\alpha}\psi_{\lambda^{-1}a}(y)
\partial_{\beta}\Delta \psi_{\lambda^{-1}a}(y)
\omega^{n-1}_{\lambda^{-1}a}(y).\eqno(54)$$
We can now use the fact that, for large $\lambda$, we have convergence of
the approximate $\delta$-function $\theta_{q\over \lambda}
\sim \delta$ to obtain
$$r_{n,\lambda^{-1}a}(x,t)=
-ne_{\alpha\beta}\partial_{\alpha}\psi_{\lambda^{-1}a}(x)
\partial_{\beta}\Delta \psi_{\lambda^{-1}a}(x)
\omega_{\lambda^{-1}a}^{n-1}(x).\eqno(55)$$
This is $q$-independent.
So we have proven that in the asymptotic regime currents are $q$
independent and represent local sources of turbulence.
\vskip .5 cm
\leftline{\bf b) The Polyakov Equation}
\vskip .5 cm
 Current densities are   composite fields. Their field theoretic
dimensions are the sum of two terms: the canonical dimension which is 2
as they are  densities, and the anomalous dimension which accounts for the
singularities appearing in short distance expansions. More precisely,
the dimension of densities is given by
$$d=2-d_{\psi_2}-d_{\psi_{n-1}}+(n+1)d_{\psi}.\eqno(56)$$
The anomalous dimension cancels all divergences appearing in the
short distance expansions of ${d\omega_{a\over \lambda}\over dt}$ and
$\omega_{a\over \lambda}^{n-1}$.
Writing $R_{n,a}(\lambda^Tt)=\int d^2 x\
\lambda^{2-d}r_{n,\lambda^{-1}a}(x,t)$, we are interested in the
asymptotic behaviour of the rescaled current densities $\tilde
r_{n,\lambda^{-1}a}(x,t)=\lambda^{2-d}r_{n,\lambda^{-1}a}(x,t)$.
 The rescaled current density is a
product of two fields:
$$\tilde r_{n,\lambda^{-1}}(x,t)= a^{d_{\psi_{n-1}}+d_{\psi_2}-2(n-1)-
(n+1)d_{\psi}}
\lambda^{2(n-1)}{\cal L}\psi_{2,\lambda^{-1}a}
\psi_{n-1,\lambda^{-1}a}\eqno(57)$$
where ${\cal L}=(L_{-2}\bar L_{-1}^2-\bar L_{-2}L_{-1}^2)$.
The first power of $\lambda$ is the remnant of the derivatives which
do not contribute to the anomalous dimension.
This product is still divergent in the continuum limit. Only a fine
tuning of the field dimensions can preserve a finite limit. Now,
suppose that the leading part of short distance expansions
at the fixed point defines an
associative algebra, i.e. the leading part $\hbox{Dom}(\psi\phi)$ of
field products  satisfies $\hbox{Dom}(\psi_1 \hbox{Dom}(\psi_2
\psi_3))
=\hbox{Dom}(
\hbox{Dom}(\psi_1\psi_2)\psi_3)$
\footnote {$^3$} {The operator algebra is always associative.
If the stream function is a $simple\  current^{[17]}$, then the leading part of
short distance expansions defines an associative algebra.}.
 This allows
$[\psi_2][\psi_{n-1}]$ and $[\psi][\psi_n]$ to be related, leading to
$$ \tilde r_{n,\lambda^{-1}a}(x,t)
=\lambda^{d_{\psi_{n-1}}+d_{\psi_2}+2(n-1)-d_{\psi_{n+1}}}
a^{d_{\psi_{n+1}}-2(n-1)-(n+1)d_{\psi}}
{\cal L} \psi_{n+1,\lambda^{-1}a}.\eqno(58)$$
The field on the right hand side is present as it is the lowest
dimensional descendent field of $\psi_{n+1}$ which turns out to be
a scalar (recall that $\psi_{n+1}$ is pseudo-scalar).
The right hand side converges provided
$$d_{\psi_{n+1}}=d_{\psi_{n-1}}+d_{\psi_2}+2(n-1).\eqno(59)$$
A solution can be easily found:
$$d_{\psi_n}={1\over 2}n(n+Q)\eqno(60)$$
Notice that $n=2$ gives the Polyakov equation. This generalised Polyakov
equation was already mentioned in ref.[11].
We have proven that the current densities  converge in
the infrared limit. The scaling properties of the fields ${\cal
L}\psi_{n+1}$ imply that currents vanish:
$$R_n=0\eqno(61)$$
This implies that the infrared behaviour of two-dimensional turbulence
is not represented by a cascade.

As a matter of fact, the Gaussian field theory with a charge $Q=-2$ at
infinity
satisfies (60). Moreover, vertex operators are simple currents;
therefore the leading part of the operator algebra is associative.
 We have found a fixed point such that fluxes of
$n$-enstrophies are finite in the infra-red limit. If the leading part
of the operator algebra
 of the fixed point defines an associative algebra, the deformed Gaussian
theory is the unique asymptotic description of 2d turbulence.
 \vskip .5 cm
\leftline{\bf V- The Coulomb Gas Picture}
\vskip .5 cm
\leftline{\bf a) The effective action}
\vskip .5 cm
The Gaussian fixed point fulfils all requirements to represent an
asymptotic state of two-dimensional turbulence. Nevertheless, we would
like to describe not only the fixed point but the effective theory in
its basin of attraction. Recall that this theory is a perturbed
conformal theory whose mass is getting smaller and smaller as one gets
closer to the fixed point.

Let us briefly sum up the time evolution of solutions of the Navier-stokes
equation.
First of all, there is a transient period when $t\le \tau$ where
solutions evolve from the initial condition to a field theory
characterised by an action $S_{\tau}$. Two possibilities are to be
envisaged.
Suppose that a solution never falls within the basin of attraction
of a conformal fixed point, then it can keep wandering in function
space never reaching any steady state. This is an unattractive case.
On the other hand, suppose that a solution falls within the basin of
attraction of a fixed point after a transient period. We will show
that the most general effective field theory in the basin of
attraction of the Gaussian fixed point with a charge at infinity
admits vortices.

Let us now describe the effective field theory in the vicinity of the
fixed point. This effective theory should
respect all the symmetries of the fixed point. We assume that
$S_{\tau}$ is in the basin of attraction of the deformed Gaussian fixed point.
Let us notice that the
deformed Gaussian fixed point is invariant under two discrete symmetries which
leave the stream function invariant:
$$\eqalign{
\phi\to -\phi,\ Q\to -Q\cr
\phi\to\phi\ +\ 2\pi\cr}
\eqno(62)$$
The first invariance ensures that $\psi$ is a real field whereas the
second leaves the vertex operator invariant.  The interaction term of
the effective theory being an even periodic function can be expanded
in a Fourier
series:
$$S_{eff}={1\over 2\pi}
\int d^2x ({1\over 2}\partial \tilde\phi_{a}^2
-\sum_{n\ge 1}z_n\cos n\tilde\phi_{a})\eqno(63)$$
where $\tilde\psi_a$ is now an effective field defined with a cut-off $a$.
The interaction term can be separated in  two terms, the first one
containing  relevant
and marginal fields, the second containing irrelevant fields. In
the infra-red regime, the deformed Gaussian theory is a
renormalisation group fixed point. It can only be perturbed by
irrelevant fields. The coupling constant of each irrelevant fields
behaves as
$z_n=z_{0n}(\lambda^{-1}a)^{{{n(n+Q)}\over 2}-2}$.
 Thus we end up with an
 effective potential:
$$V=-\sum_{n\ge n_0}z_{0n}(a\lambda^{-1})^{{{n(n+Q)}\over 2}-2}\cos
n\tilde\phi_{a}\eqno(64)$$
where $n_0$ is the minimal integer such that $n_0(n_0+Q)>4;\ i.e.\ n_0=4$.
 In the infrared regime the effective potential goes to zero.
The constants $z_{0n}$ explicitly depend on the initial probability law.
To understand the physical
meaning of this effective theory, it is convenient to draw an
 analogy with a Coulomb gas.
\vskip .5 cm
\leftline{\bf b) Turbulence as a Coulomb Gas}
\vskip .5 cm
We will show that the effective  theory is equivalent
 to a neutral Coulomb Gas$^{[16]}$.
This is a gas of charged particles allowed to appear and disappear
in a grand canonical fashion. This result stems from the expansion of the
partition function:
$$Z=\int {\cal D}\phi\  \exp (-S_{eff}),\eqno(65)$$
which can be written:
$$Z=\sum_{i\ge n_0,\ m_i,n_i=0}^{\infty}\prod_{i\ge 3}{z_i^{m_i+n_i}
\over n_i!m_i!}
\int \prod_{i}d^2x_i \prod_{i<j}({{\vert x_i -x_j\vert}\over a})^{q_iq_j},
\eqno(66)$$
where the product is not zero provided that the sum of all $q_i$ is
$-Q$.
This is nothing but the grand canonical partition function of
a Coulomb gas where charges $q_i=-\infty ...-n_0,n_0...\infty$
 interact according to
 a Coulomb potential:
$$Z=\sum_{i\ge 4\  m_i,n_i=0}^{\infty}\prod_{i\ge
n_0}{z_i^{m_i+n_i}\over{n_i!m_i!}}
\int \prod_i
d^2x_i \exp-{2\pi}[\sum_{i<j}q_i q_jV({{x_i-x_j}\over a})]\eqno(67)$$
where $V(x)=-{1\over 2\pi}\ln \vert x\vert $ is the Coulomb
potential. The gas is charged as the sum of the charges adds up to
$-Q$. Each choice of $m_i$ particles of charge $-q_i$, and $n_i$ of charge
$q_i$ gives a factorial factor
 to take into account the undistinguishability of charges.
The temperature of this gas is $1\over 2\pi$ and particles of
different charges
 have different fugacities $z_i$. Notice that the fugacities are very small.

Let us now interpret the role of these charges for turbulence.
The correlation functions of vertex operators are easily computed
in the Coulomb Gas theory. Indeed, $\exp i\phi_{a}(x) $ is represented
by an insertion of a positive charge  at point $x$. Correlation
functions of vertex operators are given by correlation functions of
charges.
Therefore, two-dimensional turbulence is equivalent to
a Coulomb gas where the probabilistic properties of the stream
function are given by the statistical behaviour
 of a positive charge.
In view of numerical simulations, our aim is to  determine
the classical configurations of the stream function and the vorticity.
 To do so, we need a further modification of the Coulomb gas
picture where the role of charges as vortices will be clear.
\vskip .5 cm
\leftline{\bf c) Two Dimensional Turbulence as a Gas of Vortices}
\vskip .5 cm
We have just
 have shown that a charge $q=1$ represents the stream function in
a probabilistic sense, i.e. correlation functions of the stream function
are equal to those of this charge in a Coulomb gas. Nevertheless,
this does not give the final description of two-dimensional turbulence
as
a gas of charges. Indeed, we are about to see that these charges
generate classical field configurations whose statistical behaviour
is exactly the one expected for a Coulomb gas.

Consider the $O(2)$ sigma model for stream function configurations,
i.e. an  action  given by
$$S_{eff}={1\over 4\pi}\int d^2x (\partial \psi)^2\eqno(68)$$
where the stream  function $\psi$ is now taken to be periodic.
Notice that this is nothing but the kinetic energy of a configuration.
As expected, the effective action can be expressed as a local action
in the velocity field.
 The stream field is probabilistically almost a free field.
Each stream function  configuration is weighted with a Gaussian
action. We allow singular stream functions:
 different singularities correspond to
 sectors in the theory, and the total partition function is obtained
summing over all sectors.
The field $\psi$ is a sum of a random part and a classical part:
$$\psi=\phi +\psi_c.\eqno(69)$$
 More precisely,  the classical field is
solution of the equation
$$\Delta \psi_c =-\sum_i q_i\delta_{x_i}\eqno(70)$$
where defects of charge $q_i$ have been considered  and $\delta$-functions
are smeared over a distance $a$. A charge $Q$ is
put at infinity, the gas is neutral if all the charges add up to $-Q$.
The classical solution reads
$$\psi_c(x)=\sum_i q_i\arctan {{(x-x_i)_2}\over {(x-x_i)_1}}.\eqno(71)$$
The integer $q_i$ is the charge of the vortex.
The random part $\phi$ will not contribute to defect correlation functions.
This decomposition tells us that a stream function configuration is entirely
specified by a countable number of vortices.

The energy of each configuration is given by:
$$S={1\over 4\pi}\int d^2x(\partial \phi+\partial \psi_c)^2.\eqno(72)$$
This can be easily computed  and gives:
$$S={1\over 4\pi}(\int d^2x (\partial \phi)^2
-4\pi \sum_{i<j} q_iq_j \ln
\vert {{x_i-x_j}\over a}\vert).\eqno(73)$$
The energy is finite if and only if:
$$\sum_i q_i +Q=0\eqno(74)$$
This is nothing but the energy of a charged  Coulomb gas.
 To complete this
correspondence, let us choose to restrict charges to
$-\infty...-n_0,n_0...\infty$.
As already shown, the stream function is represented in the Coulomb
gas picture by the insertion of a charge in the medium. It is even
simpler in the $O(2)$ sigma model. In that case, the charge $q_x$
at $x$ interacts with other free charges. The averaged response of the
medium
consists
of a classical field $\psi_c$ which depends on the fluctuating number
of free charges. The correlation function of two charges representing
the correlation function of the stream function is simply given by
the average over the fluctuating number of particles of the weight
$\exp -S(\psi_c(x,y))$ depending on the position of the two fixed charges.
This is nothing but the weight of the
classical field $\psi_c(x,y)$.
This result can be  extended to an arbitrary correlation function
. We conclude that
  these  classical configurations
represent the stream function  of a two-dimensional turbulent flow.

Finally, we can interpret  the charge $q_i$. Recall that the
vorticity is minus the Laplacian of the stream field so that for the
classical
part we have:
$$\omega_c(x)=\sum_i q_i\delta (x-x_i),\eqno(75)$$
i.e. each configuration of two-dimensional turbulence is specified by
a finite number of vortices, as shown by numerical simulations.
 The typical size of each vortex is $a$. Moreover,  these vortices are
 quantised. Notice that the
kinetic energy of a configuration is
 $-2\pi\sum_{i<j}q_i q_j \ln\vert x_i-x_j\vert $
 and the n-enstrophies are proportional to $\sum_i
q_i^n$.
In the infrared regime, the probability of charge creation is zero so
the energy as well as the $n$-enstrophies are zero.
The convergence of energy and n-enstrophies to zero is with
probability one.

We have now proven that in the vicinity of the  deformed Gaussian fixed point,
the solutions of the Navier-Stokes equations are given by a gas of
singularities where the stream function is a sum of vortices. The
non-trivial infrared limit is due  to the presence of an
integer vortex at infinity. This provides a solvable example of
two-dimensional turbulence. Moreover, the role of boundary conditions
is made explicit.

\vskip 1 cm
\leftline{\bf CONCLUSION}
\vskip 1 cm
We have shown that the asymptotic regime of two dimensional turbulence
in the basin of attraction of the deformed Gaussian fixed point is described by
a Coulomb gas. This is an interacting neutral gas filled with charges
$\vert q\vert \ge 4$.
We have explicitly shown that if the leading part of the operator
algebra of the fixed point is
associative, the deformed Gaussian fixed point is the unique
asymptotic description of 2d turbulence.
 If universality arguments can be applied to
self-organised criticality, it is clear that this must be the unique solution.

Let us comment on the comparison with numerical simulations. First of
all, it is conspicuous that the Coulomb gas has a natural description in
terms of vortices. This is indeed what simulations obtain, i.e. coherent
and stable vortices surrounded by a motionless flow. As
 initial conditions are always chosen to be
Gaussian, we expect that as soon as defects appear, solutions fall
within the basin of attraction of the Gaussian fixed point with
charges at infinity. These defects are likely to be remnants from large
initial fluctuations of the vorticity field.
It is then natural to see solutions flow rapidly along
renormalisation trajectories towards a Coulomb gas behaviour. To
support this explanation, it would be relevant to know if the infrared
energy spectrum is universal.
We predict that the infrared energy spectrum is flat, i.e.
behaves as $k^0$.
It would also be relevant
 to describe the early development of
solutions and to prove that defects appear in a finite time. This
would support our claim that the subsequent evolution of solutions is
given by a Coulomb gas.

Acknowledgements: I am grateful to P. E. Dorey for many discussions
about Conformal Field Theories. I would like to thank M. Bauer and
T. Samols for many suggestions and a very careful reading of the manuscript.

\vfill \eject
\vskip .5 cm
\centerline{\bf REFERENCES}
\vskip .5 cm
\leftline{[1] A. M. Polyakov: Nucl. Phys. {\bf B396} (1993) 367.}
\vskip .2 cm
\leftline{[2] See any textbook, e.g. ``Quantum Field Theory and
Critical Phenomena'' by J. Zinn-Justin}
 Oxford University Press
\vskip .2 cm
\leftline{[3] J.C.Williams: J. Fluid Mech. {\bf Vol 219} (1990) 361}
\vskip .1 cm
\hskip 0.2 cm M.E. Brachet, M. Meneguzzi, H. Politano and P. L. Sulem:
J. Fluid Mech. {\bf Vol 194} (1988) 333.
\vskip .2 cm
\leftline{[4] A. N. Kolmogorov: C. R. Aca. Sci. USS {\bf 243} (1941) 301.}
\vskip .1 cm
\hskip 2.2 cm J. Fluid Mech. {\bf Vol. 13} (1962) 82.
\vskip .2 cm
\leftline{[5] R. H. Kraichnan: Phys. of Fluids 10, (1967) 1417}
\vskip .2 cm
\leftline{[6] P. G. Saffmann, Stud. Appl. Math. 50 (1971) 277.}
\vskip .2 cm
\leftline{[7] H. K. Moffat in Advances in Turbulence, G. Comte- Bellot
and J. Mathieu eds. p. 284
Springer-Verlag (1986)}
\vskip .2 cm
\leftline{[8] D. Montgomery et al.: Phys. Fluid {\bf A4} (1992) 1}
\vskip .2 cm
\leftline{[9] D. Montgomery et al.: Phys. Fluid {\bf A5} (1993) 9}
\vskip .2 cm
\leftline{[10] Y. Matsuo: Mod. Phys. Lett. {\bf A8} (1993) 619}
\vskip .2 cm
\hskip 0.3 cm D. A. Lowe: Mod. Phys. Lett {\bf A8} (1993) 923.
\vskip .2 cm
\leftline{[11] G. Falkovich and A. Hannany:'' Is 2d Turbulence a
Conformal Turbulence''  hep-th 9301030}
\vskip .2 cm
\leftline{[12] G. Falkovich and V. Lebedev: ``Universal Direct Cascade
in 2d Turbulence''}
\vskip .2 cm
\leftline{[13] P. Gerard: S\'eminaire Bourbaki 44 \`eme ann\'ee,
Ast\'erisque
{\bf 206} (1992) 411}
\vskip .2 cm
\leftline{[14] J. Glimm and A. Jaffe: ``Quantum Physics'' Springer-Verlag}
\vskip .2 cm
\leftline{[15] A. B. Zamolodchichov: Sov. J. Nucl. Phys. {\bf 44}
(1986) 529.}
\vskip .2 cm
\leftline{[16] V. L. Berezinskii: Sov. Phys. JETP {\bf 32} (1970) 493.}
\vskip .2 cm
\hskip .2 cm J. M. Kosterlitz and D. J. Thouless: J. Phys. C {\bf 6}
(1973) 97
\vskip .2 cm
\hskip .2 cm B. Nienhuis in Phase Transitions and Critical Phenomena
{\bf Vol. 11} C. Domb and J. L. Lebowitz eds. Academic Press 1987.
\vskip .2 cm
\leftline{[17] A. N. Schellekens and S. Yankielowicz:
Nucl. Phys. {B334} (1990) 67.}
\end